\documentclass[final,5p,times,twocolumn]{elsarticle}
\usepackage{graphicx}
\usepackage{color}
\usepackage{amssymb}
\usepackage{amsmath}
\usepackage[nodots]{numcompress}

\journal{Stromberg Times}

\begin{document}

\begin{frontmatter}

\title{Multisite Population Epigenetic Model of Passive Demethylation}

\author[EmoryB]{S. P. Stromberg\corref{cor1}}

\date{\today}

\cortext[cor1]{Corresponding author}
\address[EmoryB]{Department of Biology, Emory University, Atlanta GA 30322}

\begin{abstract}
The methylation of DNA regulates gene expression. On cell division the methylation state of the DNA is typically inherited from parent to daughter cells. While the chemical bond between the methyl group and the DNA is very strong, changes to the methylation state do occur and are observed to occur rapidly in response to external stimulus. The loss of methylation can be active where enzyme physically breaks the bond, or passive where on cell division the newly constructed strand of DNA is not properly inherited.

Here we present a mathematical model of single locus passive demethylation for a dividing population of cells. The model describes the heterogenity in the population expected from passive mechanisms. We see that even when the site specific probabilities of passive demethylation are independent, conservation of methylation on the inherited strand gives rise to site-site correlations of the methylation state. We then extend the model to incorporate correlations between sites in the locus for demethylation rates. Biologically, correlations in demethylation rates might correspond to locus wide changes such as the inability of methyltransferase to access the locus. We also look at the effects of selection on the multicellular population. 

The model of passive demethylation not only provides a tool for measurement of parameters in loci-specific cases where passive demethylation is the dominant mechanism, but also provides a baseline in the search for active mechanisms. The model tells us that there are states of methylation inaccessible by passive mechanisms. Observation of these states constitutes evidence of active mechanisms, either de novo methylation or enzymatic demethylation. We also see that selection and passive demethylation combined can give rise to a stable heterogeneous distribution of gene methylation states in a population.
\end{abstract}


\end{frontmatter}

\section{Introduction}
The methylation of DNA is an epigenetic mechanism of gene regulation \cite{razin_dna_1984,bird_dna_2002,shoemaker_mediators_2011}. It is typically inherited from parent cell to daughter upon cell division, but there are notable epimutational divergences. These divergences give rise to tissue specific methylation patterns \cite{shoemaker_mediators_2011,deng_targeted_2009,irizarry_human_2009}, changes in gene regulation in response to stimulus \cite{youngblood_chronic_2011,bruniquel_selective_2003,martinowich_dna_2003,kangaspeska_transient_2008,metivier_cyclical_2008,kim_dna_2009}, and in aberrant cases, cancer and developmental diseases \cite{shoemaker_mediators_2011,baylin_decade_2011,doerfler_dna_2010}. Understanding the population dynamics of these epimutations is thus critical to understanding a number of important biological processes.

In vertebrates, methylation occurs primarily on the cytosines of CpG dyads, where a carbon-carbon bond connects the methyl group to the cytosine. These sites are symmetric such that a methylation site on the Watson strand will have a counterpart on the reverse compliment Crick strand. 

On cell division, daughter cells will have a methylated parent strand and a newly constructed daughter strand which is initially unmethylated. Inheritance of the methylation state of the parent is performed by maintenance methylation where the enzyme DNMT1, finding a site on the parent strand that is hemimethylated (one cystine in the dyad methylated, the other unmethylated), adds a methyl group to the corresponding site on the daughter strand. 

Epimutations are classified as either passive or active. In passive epimutations maintenance methylation does not occur leaving the dyad hemimethylated. In active epimutations either a methyltransferase adds a methyl group (\emph{de novo} methylation) or enzymatic activity leaves an unmethylated cytosine where a methylated one had been (active demethylation) \cite{wu_active_2010,ooi_colorful_2008}. The loss of a methyl group without enzymatic action is thermodynamically unlikely due to the strength of the carbon-carbon bond and the ability of enzymes to break this bond is in dispute \cite{wu_active_2010,ooi_colorful_2008}. Active demethylation is typically assumed to take place via base excision and repair. 

Previous population epigenetic models of methylation dynamics have looked at maintaining tissue specific methylation patterns \cite{pfeifer_polymerase_1990,otto_dna_1990,genereux_population-epigenetic_2005,fu_statistical_2010}. These models have made the approximation that the methylation status at each site is independent of the other sites. In maintenance methylation this approximation is useful. In this manuscript we show that for loci-specific demethylation occurring in response to external stimuli, the site independent approximation omits important features.

In this manuscript we model the population epigenetics of passive demethylation.  We study a dividing population of cells with a multisite locus where passive demethylation can occur. The model keeps track of site-site correlations which result from the initially methylated state and the strength of the carbon-carbon bond. The result of this is a heterogeneous population of cells with a distribution of methylation patterns. The model is extended to incorporate other sources of site-site correlations and to include selection. 

The model quite notably omits \emph{de novo} methylation and active demethylation. \emph{De novo} methylation is estimated to occur 10 to 50 times less frequently than maintenance methylation \cite{genereux_population-epigenetic_2005} and its omission is a simplifying approximation. While there is evidence of active demethylation \cite{bruniquel_selective_2003}, the mechanisms are not yet well understood. The model in this paper thus also serves as a baseline such that deviations from it, suggest signatures of active mechanisms.

\section{Model}
Here we construct a multicellular model of a single locus with multiple methylation sites. We neglect recombination as all sites are assumed to be close enough that it is unlikely to occur. The model keeps track of the fractions of cells in all possible methylation states and examines how these fractions change with the number of cell divisions. We first construct a single site model. The single site model is equivalent to those previously studied \cite{pfeifer_polymerase_1990,otto_dna_1990,genereux_population-epigenetic_2005,fu_statistical_2010}. We then construct multisite models illustrating the inaccuracy of site-site independence in the dynamic case. We then extend the model to include not only the site-site correlations in state, but where the error rates on the sites are correlated as well. We also extend the model to examine the effects of selection on the population.

For a single site (CpG dyad) there are four possible states: methylated, hemimethylated Watson polarity, hemimethylated Crick polarity, and demethylated. We use the following symbols to illustrate these states:
\begin{figure}[htbp]
	\centering
		\includegraphics[scale=1]{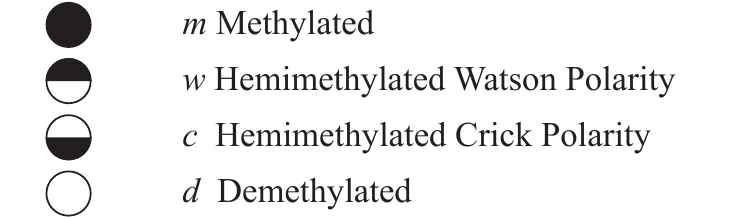}
	\label{fig:label}
\end{figure}

Our model considers the population after maintenance methylation has occurred. It is typically assumed that maintenance methylation occurs immediately following DNA replication \cite{leonhardt_targeting_1992}. This model therefore describes cells in the G0 or G1 phases.

Without demethylase, after division the methyl groups on the inherited parent strand remain. Maintenance methylation occurs at sites on the newly constructed strand if the site on the parent strand is methylated. Occasionally maintenance methylation fails to methylate the site on the newly constructed daughter strand leaving the site in a hemimethylated state, (Watson polarity if the parent strand was the Watson strand, and Crick polarity if the parent strand was the Crick strand). We refer to this omission as an error, though it may be biologically advantageous for this to occur, and label the probability of an error occurring at a single site $\mu$. This sets the probability of maintenance methylation occurring to have rate $1-\mu$.

\begin{figure}[htbp]
	\centering
		\includegraphics[scale=1]{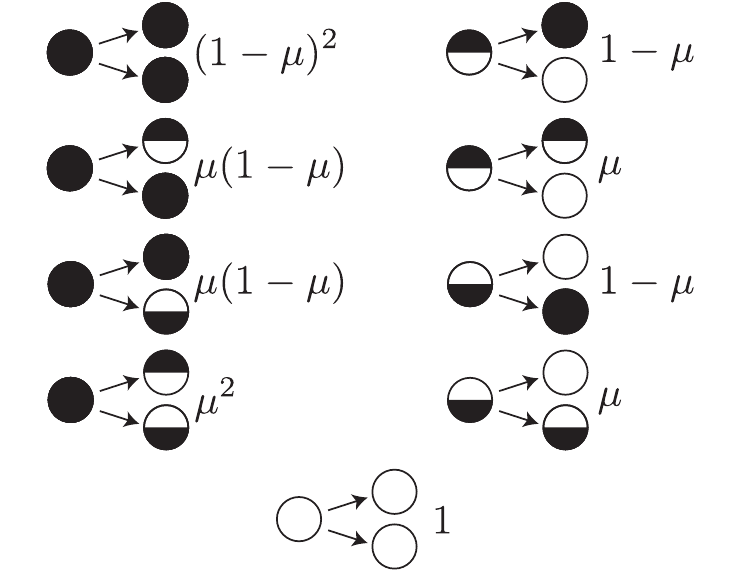}
	\caption{Possible transitions on cell division with probability of each outcome, where the probability of a daughter having a hemimethylated state is given by $\mu$ and having a methylated state is $1-\mu$. We also see that the totally methylated state (bottom) is trapping having no transitions to other states.}
	\label{fig:Transitions}
\end{figure}

The possible transitions for a single site from a parent to the pair of daughters are shown in Fig.~\ref{fig:Transitions} along with the probabilities of each outcome.  We see that if the parent is in the methylated state there can be no daughter in the demethylated state, that a parent in either polarity hemimethylated state will give a single demethylated daughter cell, and that demethylated sites are trapping, having only daughters in the demethylated state. 

Multisite dynamics will require us to analyze the inheritance of the Watson and Crick strands from the parent separately, though the dynamics are symmetric. For the single site, taking the daughters from the top on each division in Fig.~\ref{fig:Transitions} for the Watson strand and those daughters on the bottom for the Crick strand we segregate as in Fig.~\ref{fig:Coding}:
\begin{figure}[htbp]
	\centering
		\includegraphics[scale=1]{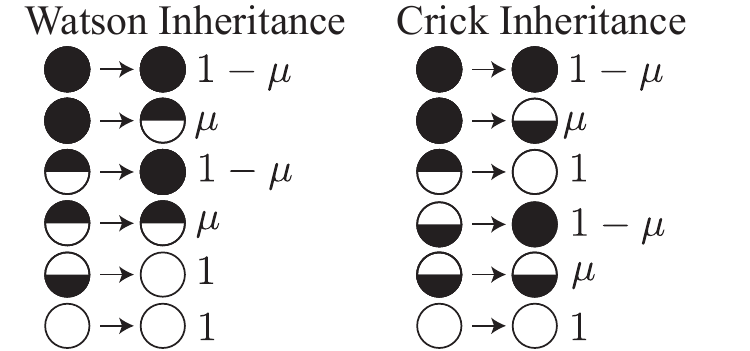}
	\caption{The transitions with fractional rates for the daughters inheriting the Watson strand and those inheriting the Crick strand.}
	\label{fig:Coding}
\end{figure}
 
From these transitions we can construct transition matrices. With the states labeled as basis vectors according to: 
\begin{eqnarray}
	m=\begin{bmatrix}
	1\\
	0\\
	0\\
	0
	\end{bmatrix},~~~	w=\begin{bmatrix}
	0\\
	1\\
	0\\
	0
	\end{bmatrix},~~~	c=\begin{bmatrix}
	0\\
	0\\
	1\\
	0
	\end{bmatrix},~~~	d=\begin{bmatrix}
	0\\
	0\\
	0\\
	1
	\end{bmatrix},
\end{eqnarray}
the above transitions give us the Watson strand transition matrix:
\begin{eqnarray}
	\boldsymbol{W} &=& \begin{bmatrix}
	1-\mu 			& 1-\mu 	& 0 	& 0\\
	 \mu	&  \mu		& 0						& 0\\
	0 	& 0						& 0		& 0\\
	0				& 0		& 1			& 1
	\end{bmatrix},
\end{eqnarray}
and the Crick strand transition matrix:
\begin{eqnarray}
	\boldsymbol{C} &=& \begin{bmatrix}
	1-\mu 			& 0 	& 1-\mu 	& 0\\
	 0	&  0		& 0						& 0\\
	\mu 	& 0						& \mu		& 0\\
	0				& 1		& 0			& 1
	\end{bmatrix}.
\end{eqnarray}

\subsection{One Site Model}
We now use the transition matrices to construct a probabilistic model for a single site. 
If we describe the fractions of cells with each methylation state at generation $g$ as a vector $\vec{P}^{(1)}(g)$, e.g.:
\begin{equation}
	\textrm{Pr}(m|g)=m\cdot\vec{P}^{(1)}(g),
\end{equation}
gives the fraction of cells that are methylated, the vector at $g+1$ is given by:
\begin{eqnarray}
	\vec{P}^{(1)}(g+1) &=& \frac{1}{2}\left[\boldsymbol{W} + \boldsymbol{C} \right]\vec{P}^{(1)}(g)\\
	&=& \begin{bmatrix}
	1-\mu 			& \frac{1}{2} (1-\mu) 	& \frac{1}{2} (1-\mu) 	& 0\\
	\frac{1}{2} \mu	& \frac{1}{2} \mu		& 0						& 0\\
	\frac{1}{2} \mu	& 0						& \frac{1}{2} \mu		& 0\\
	0				& \frac{1}{2}			& \frac{1}{2}			& 1
	\end{bmatrix} \vec{P}^{(1)}(g)~~~~~\label{eq:A}\\
	\vec{P}^{(1)}(g+1) &=& \boldsymbol{T}^{(1)}~ \vec{P}^{(1)}(g),\label{eq:B}
\end{eqnarray}
where we have added superscripts to the transition matrix and state vector to indicate that it is for a single site model. The model assumes that at each generation every cell divides and that there is zero cell death. 

\begin{figure}[htbp]
	\centering
		\includegraphics[scale=1]{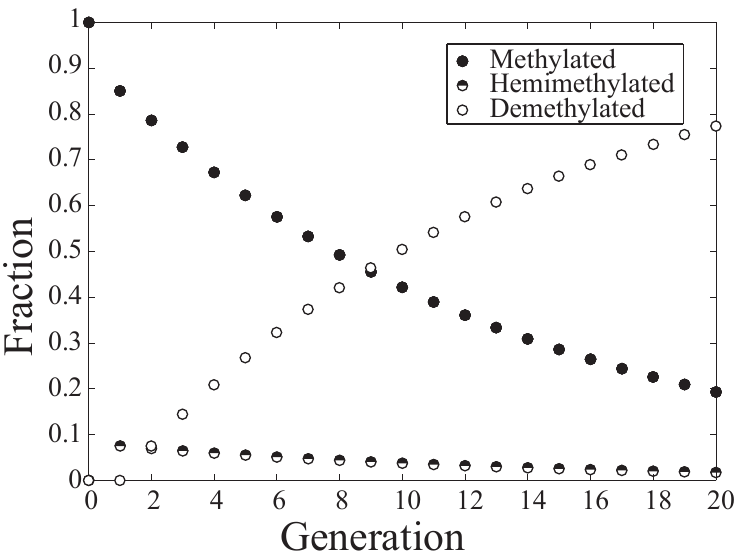}
	\caption{The fractions of sites with each methylation state as a function of generation number with $\mu=0.15$ and initially all sites in the population methylated. Only one of the two hemimethylated states is shown as the probabilities of the two hemimethylated states are equivalent.}
	\label{fig:OneProb}
\end{figure}

Starting with all cells methylated ($\vec{P}(0) =m$) and taking $\mu=0.15$ the model generates the dynamics shown in Fig.~\ref{fig:OneProb}.
We see from this figure that the fraction of methylated sites declines with generation number. This is due to the exponential growth in population size. We also see that the time-scale of the transition from methylated to demethylated is set by $1/\mu$.

\subsection{Multisite Models}
Here we consider a single gene with multiple sites that can be methylated. We do not consider recombination as the rate of recombination on a single gene is too low. We describe an $n$ site sequence using the Kronecker product of the single site state vectors. Given a sequence $\{s_1,s_2,\ldots,s_n\}$ where each $s_i$ is one of the four states $s_i=m,w,c$ or $d$ corresponding to the states described above, the vector describing the sequence is given by:
\begin{equation}
	\{s_1,s_2,\ldots,s_n\} = s_1 \otimes s_2 \otimes \ldots \otimes s_n,
\end{equation}
and we use the curly braces above throughout the text as an abbreviation of the Kronecker products.

The Kronecker product is defined for two vectors of length 4 according to:
\begin{equation}
	\{a,b\} = \begin{bmatrix}
	a_1 b_1\\
	a_1 b_2\\
	\vdots\\
	a_2 b_1\\
	a_2 b_2\\
	\vdots\\
	a_4 b_4
	\end{bmatrix},
\end{equation}
a vector of length 16. Similarly for two $4\times 4$ matrices the Kronecker product generates a $16 \times 16$ matrix defined by:
\begin{equation}
	\{\boldsymbol{A} , \boldsymbol{B}\} = \begin{bmatrix}
	A_{11}\boldsymbol{B} & \ldots & A_{14}\boldsymbol{B}\\
	\vdots & \ddots & \vdots\\
	A_{41}\boldsymbol{B} & \ldots & A_{44}\boldsymbol{B}
	\end{bmatrix}.
\end{equation}
As an example, the two site sequence $\{m,m\}$ is given by $(1,0,\ldots,0)$, $\{m,c\}$ is given by $(0,1,0,\ldots,0)$ and so on.

Using this formalism the probability of a multisite sequence with $n$ sites, at generation $g$ being in the state described by $\{s_1,s_2,\ldots,s_n\}$ is given by:
\begin{equation}
	\textrm{Pr}(\{s_1,s_2,\ldots,s_n\}|g) = \{s_1,s_2,\ldots,s_n\} \cdot \vec{P}^{(n)}(g).\label{eq:D}
\end{equation}
We can also see that the multisite probability vector is given by the Kronecker product of the individual site probability vectors:
\begin{equation}
	\vec{P}^{(n)}(g) = \{\vec{P}^{(1)}_1(g),\vec{P}^{(1)}_2(g),\dots,\vec{P}^{(1)}_n(g)\}\label{eq:C}
\end{equation}
The dynamics of the system are now obtained with a transition matrix $\boldsymbol{T}^{(n)}$ that given $\vec{P}^{(n)}(g)$ returns $\vec{P}^{(n)}(g+1)$.

\subsubsection{ID Omits Essential Features of Passive Demethylation}
Previous multisite models of methylation dynamics have studied large scale maintenance of methylation patterns \cite{pfeifer_polymerase_1990,otto_dna_1990,genereux_population-epigenetic_2005,fu_statistical_2010}. These models have used an approximation where the probabilities of different states at each site are independently distributed (ID). This approximation has been useful in the study of maintenance, but here we show that even when the probability of error is independent at each site, the ID approximation omits important features in the system which result from the conservation of carbon-carbon bonds. In dynamic scenarios of passive demethylation the conservation of these bonds gives rise to heterogeneity in the population, site-site correlations, and the absence of opposite polarity hemimethylated sites.

The ID approximation for multisite models is achieved by taking the Kronecker product of the one site transfer matrix $\boldsymbol{T}^{(1)}$ given in Eq.~\ref{eq:B}.
\begin{equation}
	\boldsymbol{T}^{(n)} \approx \{\boldsymbol{T}_1^{(1)},\boldsymbol{T}_2^{(1)},\dots,\boldsymbol{T}_n^{(1)}\}.\label{eq:H}
\end{equation}
Taken with Eq.~\ref{eq:B} and \ref{eq:C} we have:
\begin{eqnarray}
	\vec{P}^{(n)}(g+1) &\approx& \boldsymbol{T}^{(n)}~ \vec{P}^{(n)}(g)\label{eq:K}\\
	&=& \{\boldsymbol{T}_1^{(1)},\dots,\boldsymbol{T}_n^{(1)}\}\cdot\{\vec{P}^{(1)}_1(g),\dots,\vec{P}^{(1)}_n(g)\}\\
	&=&\{\boldsymbol{T}_1^{(1)}\vec{P}^{(1)}_1(g),\dots,\boldsymbol{T}_n^{(1)}\vec{P}^{(1)}_n(g)\}\label{eq:F}\\
	&=&\{\vec{P}^{(1)}_1(g+1),\dots,\vec{P}^{(1)}_n(g+1)\},
\end{eqnarray}
where in Eq.~\ref{eq:F} we have used the mixed product property of the Kronecker product (i.e. $(\boldsymbol{A}\otimes\boldsymbol{B})(\boldsymbol{C}\otimes\boldsymbol{D}) = \boldsymbol{A}\boldsymbol{C}\otimes \boldsymbol{B}\boldsymbol{D}$). The probability of a specific sequence is then given by Eq.~\ref{eq:D}:
\begin{eqnarray}
	Pr(\{s_1,\ldots\,s_n\}|g+1) &\approx& \{s_1,\ldots,s_n\} \cdot \vec{P}^{(n)}(g+1)\\
	&=&\prod_{i=1}^n s_i \cdot \vec{P}_i^{(1)}(g+1),\label{eq:E}
\end{eqnarray}
which is a more common notation for an ID approximation.

The ID approximation neglects correlations between sites that are the result of bond conservation. As the methylation status of a site is a categorical variable there is no simple summary statistic to illustrate the effect of the maintenance correlations. 

\begin{figure}[htbp]
	\centering
		\includegraphics[scale=1]{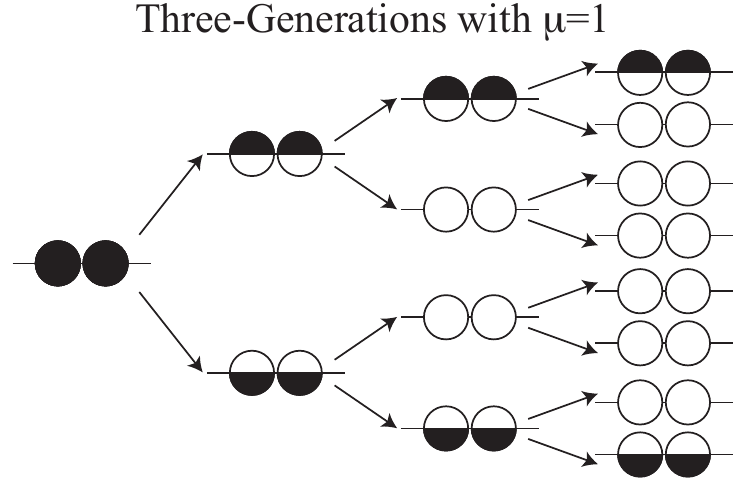}
	\caption{A limiting example to underscore the importance of correlations between sites. Because the carbon-carbon bonds are not expected to be broken in passive demethylation the parent strands are maintained in the population (shown at edges of population). This maintenance is responsible for site-site correlations even when errors occur with independent probabilities. With $\mu<1$ these strands are still maintained and serve as a continual source of cells with high numbers of methyl groups. The conservation of these strands constitutes asymmetric cell division. We also see there are no states with a combination of opposite polarity hemimethylated sites ($w$ and $c$) as is predicted by independently distributed models.}
	\label{fig:Mu1Example}
\end{figure}

In the absence of a summary statistic for categorical variables consider the dynamic in Fig.~\ref{fig:Mu1Example}. Here we illustrate the first 3 divisions of a system with the limiting case of $\mu=1$ for both sites, starting with $\{s_1,s_2\}=\{m,m\}$. This limiting example is deterministic and the first division results in two hemimethylated strands ($\{w,w\}$ and $\{c,c\}$). Subsequent divisions maintain the hemimethylated strands while adding demethylated strands. 

We see in this example that if a methyl group is detected at the first site in the locus it predicts with certainty that the second site will also be methylated. Bond conservation is the source of these site-site correlations, and constitutes an asymmetric cell division.

For a single site with $\mu=1$, starting with $\vec{P}(0)=m$ repeated application of Eq.~\ref{eq:A} gives us $\vec{P}(3)=(0,1/8,1/8,3/4)$ in correspondence with the single sites illustrated in Fig.~\ref{fig:Mu1Example}. 

For multiple sites one can quickly see that the ID approximation in Eq.~\ref{eq:E} yields poor correspondence with what is illustrated. For example the ID approximation yields a probability of $9/16$ in this case for observing the sequence $\{d,d\}$ in the third generation, the probability of which is easily seen from the figure to be $3/4$. This underestimate on Pr$(\{d,d\}|3)$is balanced by overestimating the probabilities for unobserved states such as $\{c,d\}$.

\subsubsection{Exact Model}
To analyze passive demethylation without using the ID approximation one must consider the transfer matrices for the Watson and Crick $\boldsymbol{W}$ and $\boldsymbol{C}$ strands separately. Performing a multi-site analysis analogous to the one in Fig.~\ref{fig:Coding}, we find that the $n$ site Watson strand transition matrix $\boldsymbol{W}^{(n)}$ is given by:
\begin{equation}
	\boldsymbol{W}^{(n)} = \{\boldsymbol{W}_1,\boldsymbol{W}_2,\ldots, \boldsymbol{W}_n\},
\end{equation}
and similarly:
\begin{equation}
	\boldsymbol{C}^{(n)} = \{\boldsymbol{C}_1,\boldsymbol{C}_2,\ldots, \boldsymbol{C}_n\}.
\end{equation}
We note here that the $\mu$ values for the different sites do not need to be equivalent in constructing these matrices. Combining these two matrices gives us the exact multisite transfer matrix:
\begin{equation}
	\boldsymbol{T}^{(n)} = \frac{1}{2}[\boldsymbol{W}^{(n)}+\boldsymbol{C}^{(n)}].\label{eq:G}
\end{equation}
For the $\mu=1$ case illustrated in Fig.~\ref{fig:Mu1Example} Eq.~\ref{eq:G} reproduces the correct fractions as well as in the general case. 

Algebraically, the difference between the expressions in Eq.~\ref{eq:G} and \ref{eq:H} can be found using the identities of the Kronecker product $\{\boldsymbol{A}, (\boldsymbol{B}+\boldsymbol{C})\} = \{\boldsymbol{A},\boldsymbol{B}\}+ \{\boldsymbol{A},\boldsymbol{C}\}$, and $\{(\boldsymbol{A}+ \boldsymbol{B}),\boldsymbol{C}\} = \{\boldsymbol{A},\boldsymbol{C}\}+ \{\boldsymbol{B},\boldsymbol{C}\}$. We find for the $\boldsymbol{T}^{(2)}$ case generated by Eq.~\ref{eq:H}:
\begin{eqnarray}
	\{\boldsymbol{T}^{(1)},\boldsymbol{T}^{(1)}\}&=&\frac{1}{4}\{(\boldsymbol{W}+ \boldsymbol{C}),(\boldsymbol{W}+ \boldsymbol{C})\}\\
	&=&\frac{1}{4}\left[\{\boldsymbol{W},\boldsymbol{W}\} + \{\boldsymbol{C},\boldsymbol{C}\} + \{\boldsymbol{C},\boldsymbol{W}\} + \{\boldsymbol{W},\boldsymbol{C}\} \right]~~~~~~~\\
	&=&\frac{1}{2}\boldsymbol{T}^{(2)}+\frac{1}{4}\left[\{\boldsymbol{C},\boldsymbol{W}\} + \{\boldsymbol{W},\boldsymbol{C}\} \right].
\end{eqnarray}
We therefore see that there are cross-terms in the ID approximation not found in the exact model.

\subsection{Correlated Errors}
Passive demethylation is a result of the maintenance methylation enzymes failing to correctly methylate the corresponding state on the newly constructed strand of DNA. We may therefore suspect that if this enzyme is in low quantity or otherwise blocked from accessing the locus, that the chance of error at all sites in the locus is increased. This generates additional correlations between sites. 

To model this effect we consider multiple possible values of $\mu$ each denoted with a subscript $(\mu_1,\mu_2,\ldots)$. When a cell divides there is probability given by $p(\mu_j)$ that the daughter cells are generated with error rate at every site in the locus equal to $\mu_j$. We also have the expected value of $\mu$ given by:
\begin{equation}
	\overline{\mu} = \sum_{j=1}^m p(\mu_j) \mu_j.
\end{equation}

The quantity $p(\mu_j)$ equivalently denotes the expected fraction of cells to divide with the error rate $\mu_j$. In the case of all sites in a dividing cell having equivalent error rates, this allows us to write down the transfer matrix as the weighted sum of transfer matrices:
\begin{equation}
	\boldsymbol{T}^{(n)} = \sum_{j=1} p(\mu_j) \boldsymbol{T}^{(n)}(\mu_j),\label{eq:I}
\end{equation}
where we have made the individual transfer matrices an explicit function of the error rates. For a continuous distribution of $\mu$ values we can extend the sum to an integral:
\begin{equation}
	\boldsymbol{T}^{(n)} = \int_{0}^\infty p(\mu) \boldsymbol{T}^{(n)}(\mu) d\mu.
\end{equation}

When the methylation error rates differ from site to site, the model is more complex but can be constructed in a similar manner. One would have a probability of a set of error rates and take the analogous sum over all sets considered.

\subsection{Selection}
Methylation regulates gene expression and in many cases this can affect cell survival. This will bias the distribution of cells. To incorporate selection we can multiply the transition matrix by a diagonal matrix $\boldsymbol{S}^{(n)}$ such that:
\begin{equation}
	\{s_1,s_2,\ldots,s_n\} \cdot  \boldsymbol{S}^{(n)},
\end{equation}
gives the survival fraction for cells with sequence $\{s_1,s_2,\ldots,s_n\}$.

Multiplication by the matrix $\boldsymbol{S}^{(n)}$ leaves the probability vector un-normalized, such that the entries no longer sum to one. In order keep the probability vector normalized when selection is present we must divide by the sum of the elements of $\vec{P}^{(n)}(g+1)$. This gives us the equation:
\begin{equation}
	\vec{P}^{(n)}(g+1) = \frac{1}{\left|\boldsymbol{S}^{(n)} \vec{P}^{(n)}(g) \right|}  \boldsymbol{T}^{(n)} \boldsymbol{S}^{(n)} \vec{P}^{(n)}(g),\label{eq:J}
\end{equation}
where the bars in the pre-factor indicate taking the norm of the resulting vector.

Eq.~\ref{eq:J} describes selection that occurs before division. In general $\boldsymbol{S}^{(n)}$ and $\boldsymbol{T}^{(n)}$ do not commute. If we are instead concerned with selection that occurs after division we are required to modify the equation such that:
\begin{equation}
	\vec{P}^{(n)}(g+1) = \frac{1}{\left|\boldsymbol{S}^{(n)} \boldsymbol{T}^{(n)} \vec{P}^{(n)}(g) \right|}  \boldsymbol{S}^{(n)} \boldsymbol{T}^{(n)} \vec{P}^{(n)}(g).
\end{equation}

In general the order of operations affects the interpretation of data.


\section{Results}
\begin{figure}[htbp]
	\centering
		\includegraphics[scale=1]{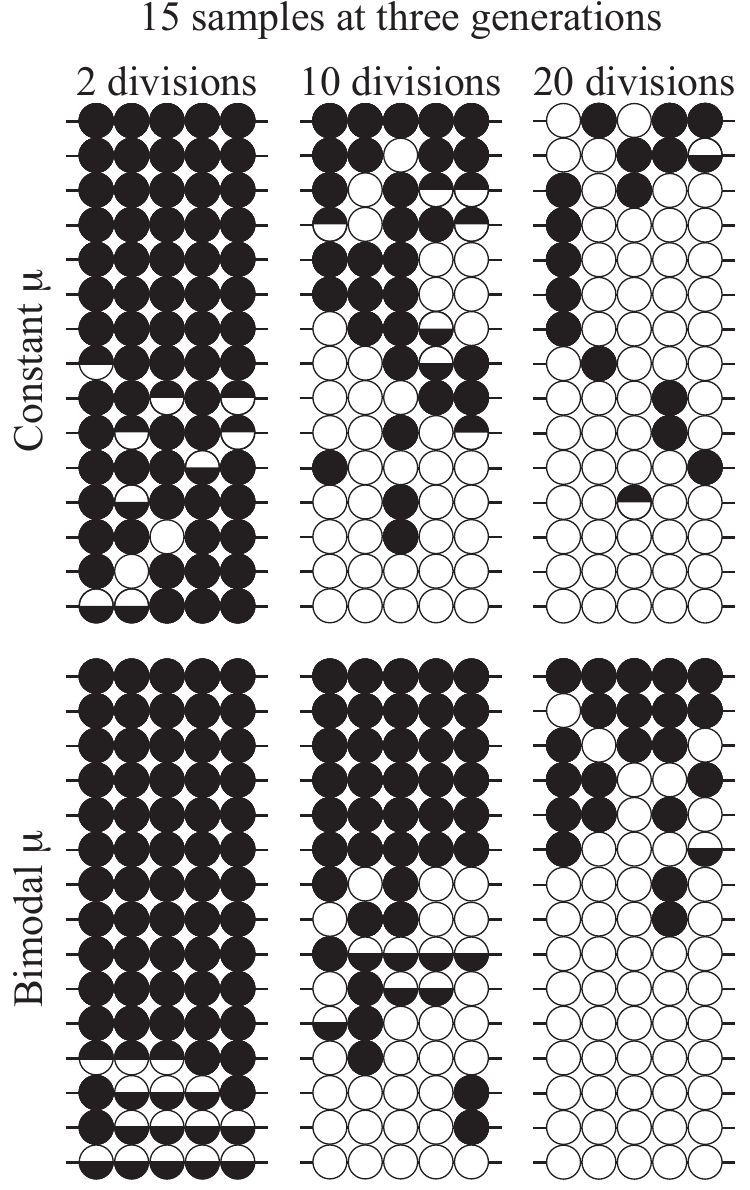}
	\caption{15 samples of a five site model taken at three different generations using Eq.~\ref{eq:G} (top) and Eq.~\ref{eq:I} (bottom). The samples are reordered in the columns to guide the eye.  In the top samples, all cells divide with the site error rate of $\mu=0.15$, while at the bottom 3/4 of cells divide with $\mu=0$ and 1/4 with $\mu=0.6$ preserving $\overline{\mu}=0.15$. The correlated errors are most apparent in the clustering observed after 10 divisions. We also see that there are no samples containing opposite polarity hemimethylated states.}
	\label{fig:Simulation}
\end{figure}

Fig.~\ref{fig:Simulation} illustrates the typical behavior of the models given by Eq.~\ref{eq:G} (top) and Eq.~\ref{eq:I} (bottom). Here we have drawn 15 samples from the distributions after 2, 10 and 20 divisions. We started the simulation with all sites in all cells being methylated. In the top samples, all cells divide with error rate $\mu=0.15$, while in the bottom we have 3/4 of cells dividing with $\mu=0$ and 1/4 of cells dividing with $\mu=0.6$ thus maintaining the same $\overline{\mu}$ as in the top. The correlated errors are most apparent in the clustering of the methyl groups in the samples. 

\subsection{Inaccessible States}
In Fig.~\ref{fig:Simulation} we see that there are no samples having opposite polarity hemimethylated states (i.e.~no states with both Watson and Crick hemimethylated sites). These states are strictly disallowed by passive demethylation. 

\begin{figure}[htbp]
	\centering
		\includegraphics[scale=1]{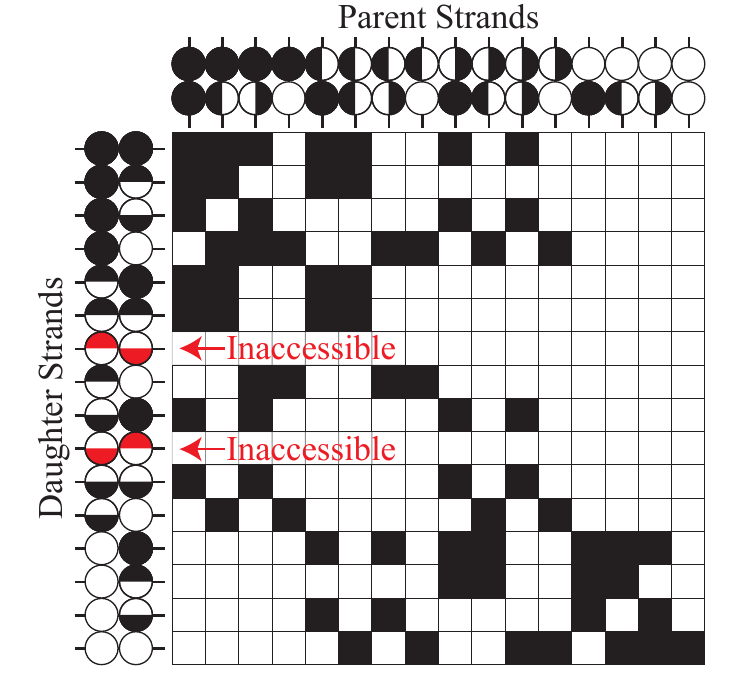}
	\caption{Illustration of the matrix of allowed transitions (filled squares) from parent (column) to daughter (row) given by Eq.~\ref{eq:G} for two sites with $0<\mu<1$. We see that the state with both sites demethylated feeds only into itself, but more notably there are two states that are inaccessible by passive means. In $n$ site models any state with opposite polarity hemimethylated sites is disallowed. Observation of such states is evidence of active mechanisms.}
	\label{fig:Matrix}
\end{figure}

To illustrate the origin of this effect in the model, Fig.~\ref{fig:Matrix} shows the matrix of allowed transitions for $0<\mu<1$. Here, for a two site model, black boxes indicate a possible transition from the parent column to the daughter row. From this figure we see that there are two states that are inaccessible having opposite polarity hemimethylated sites. No parent state is capable of generating either of these states upon division, not even these states themselves. 

The ID model yielded finite probabilities for these states, but from the full model in Eq.~\ref{eq:G} we see that no passive demethylation can result in these states. With higher numbers of sites as in Fig.~\ref{fig:Simulation}, any state with opposite polarity hemimethylated sites is strictly inaccessible by passive demethylation.

\subsection{Summary Distributions}
For the full model we consider all possible states. At four states per site and five sites this gives $4^5 = 1024$ states. As a summary of the features of the model we consider the total number of methyl groups on the Watson strand of each cell. As the dynamics of the Watson and Crick strands are symmetric in the model the sum of methyl groups on the Crick strand will have an equivalent distribution when there is no selection.

\begin{figure}[htbp]
	\centering
		\includegraphics[scale=1]{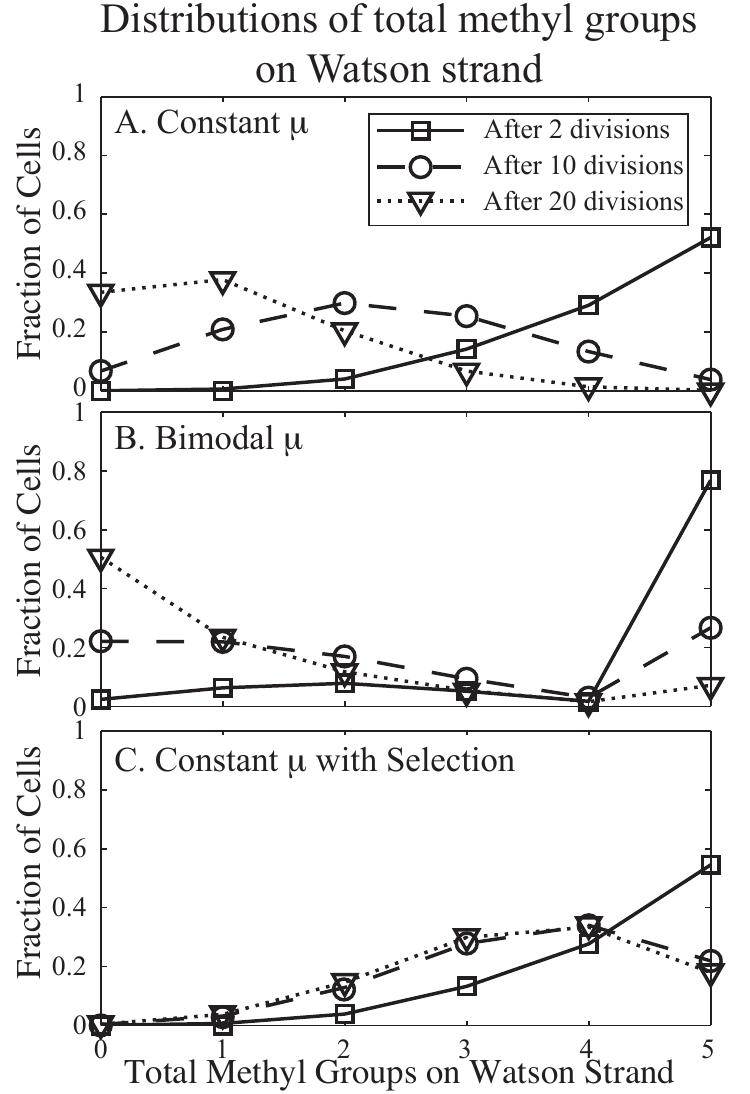}
	\caption{The distribution of methyl group numbers on the Watson strands after 2 divisions, 10 divisions, and 20 divisions. A shows distributions when all sites in all cells have $\mu=0.15$. B maintains $\overline{\mu}=0.15$ but most cells have no error and a quarter of the cells divide with a high error rate of $\mu=0.6$. C shows the case with constant $\mu$ and selection against demethylated cells. We see in C that the distribution is stable having changed very little from 10 divisions to 20 divisions. In all simulations the initial fraction of cells with all five sites methylated is 1 (not shown). Without selection the distributions tend to low methyl group numbers. The bimodal $\mu$ distribution model is characterized by fewer cells with intermediate numbers of methyl groups.}
	\label{fig:distributions}
\end{figure}

For an $n$ site system the reduced representation has $n+1$ states ranging from zero methyl groups to $n$ methyl groups. Fig.~\ref{fig:distributions} (A) shows the probability distribution of cells for $n=5$ as a function of the total number of methyl groups on the Watson strand after 2 divisions, 10 divisions, and 20 divisions with $\mu=0.15$. 

In Fig.~\ref{fig:distributions} (B) we see the effects of correlated errors (Eq.~\ref{eq:I}) on this distribution. This distribution like the bottom samples in Fig.~\ref{fig:Simulation} is bimodal, having 3/4 of cells dividing with $\mu_1=0$ and 1/4 dividing with $\mu_2=0.6$, preserving $\overline{\mu}=0.15$. Here we see that the distribution after two divisions has a marked divergence from the independent errors model, having very few cells with a single loss of methylation on the Watson strand. After 10 divisions the distribution is bimodal in comparison to the single peaked distribution from the independent errors model.

\subsection{Stable Distributions}
Fig.~\ref{fig:Mu1Example} illustrates the conservation of methylation on the grandmother strands when $\mu=1$. The grandmother strands are those that where present in the population when expansion began which we assume to be completely methylated. Without enzymatic action to remove the methyl group, these strands remain methylated indefinitely. When $\mu<1$, as long as the cells with these strands survive, the grandmother strands act to seed the population with new methylated cells.

The demethylated state is trapping. Once a site is demethylated in this model, all descendants will also be demethylated. This acts to dilute the effects of the grandmother cells seeding the population. With selection however a stable heterogeneous distribution can be achieved. 

As an example of this we use Eq.~\ref{eq:J} and construct a model such that cell fitness is a linear function of the number of methyl groups on the Watson strand. In this model, states with 5 methyl groups on the Watson side have $\boldsymbol{S}^{(n)}_{ii}$ = 1. With 4 methyl groups, $\boldsymbol{S}^{(n)}_{ii}$ = 4/5, and so on. This model is not symmetric between the Watson and Crick strands as we are considering selection only acting on the state of the Watson strand.  

Stable distributions are found by computing the eigenvectors of the matrix given by $\boldsymbol{T}^{(n)}\boldsymbol{S}^{(n)}$. There are in general $4^n$ eigenvectors for this matrix. Eigenvectors with negative valued components aren't biologically meaningful, and if the error rates ($\mu$) are equivalent at multiple sites, then there is significant degeneracy. Here we are interested in the eigenvector associated with the dominant eigenvalue which a population starting completely methylated will converge to.

In Fig.~\ref{fig:distributions}(C) we show the summary distribution for the model with selection. Here we see very little change in the shape of the distribution from 10 divisions to 20 divisions. This illustrates that survival of the grandmother strand can seed the population and maintain a constant and heterogeneous distribution of cells. 

\section{Discussion}
All stochastic systems beginning in the same state will be initially correlated (autocorrelation) but this effect will decay as the two sites diverge. Models of demethylation using site independent distributions starting with all sites methylated also have this type of correlation. 

We have seen that the maintenance of carbon-carbon bonds gives rise to an additional type of correlation that is important in the dynamics of passive demethylation. While the methylation status of a site is categorical and thus has no standard summary statistic for analyzing the correlations between sites, we see that from the extreme example in Fig.~\ref{fig:Mu1Example} that observation of a methyl group on the first site predicts with certainty that there will be one on the second site. As the error rate for maintenance methylation decreases the effect is still present but less strong.

Additionally we see from Fig.~\ref{fig:Matrix} the inaccessibility by passive demethylation of opposite polarity hemimethylated states. This strict site-site anticorrelation is present for all error rates. Site independent distributions (ID) fail to capture this effect. Opposite polarity hemimethylated sites have been observed \cite{laird_hairpin-bisulfite_2004} and the observation was previously used to infer that \emph{de novo} methylation could result in a hemimethylated state \cite{genereux_population-epigenetic_2005}. Depending on the specific mechanisms of the process this may also be a possible outcome of active demethylation. The detection of hemithylated states requires an additional step in sequencing \cite{miner_molecular_2004,laird_hairpin-bisulfite_2004}, but when questions are present of active versus passive mechanisms arise this extra step can might help.

These correlations by bond maintenance are an example of asymmetric cell division. As there are no opposite polarity hemimethylated dyads, if a parent having only Watson polarity hemimethylated sites divides, the daughter receiving the Crick strand will be unmethylated at each site that was hemimethylated in the parent. The daughter inheriting the Watson strand will either be methylated at these sites or hemimethylated. This asymmetry between daughters generates heterogeneity in the population. This heterogeneity is explored in Fig.~\ref{fig:distributions}(A). 

In combination with selection, the heterogeneity resulting from asymmetric cell division can give rise to a stable distribution (Fig.~\ref{fig:distributions}C). This might correspond to heterogeneity found in homeostatic conditions. The model then allows one to make connections between dynamics and cross-sectional data. Observation of a heterogeneous distribution of methylation states might also be used as a biomarker for the stimulus that triggered demethylation.

The simulations in this paper have all assumed that cells are initially methylated and begin to loose methylation upon the first division with a constant rate $\mu$ for the remainder of the simulation. This corresponds to a scenario where an external stimulus begins at the start of the simulation and has a constant effect throughout. An alternate method might be to consider time dependent parameters. The exact model in this paper can easily be altered for such scenarios.

Passive demethylation could either be a global process where the entire cell is deficient in the maintenance methyltransferase or be a local effect where a particular locus is partially or completely sequestered from the maintenance enzymes. In the global case we would expect that the entire genome would be loosing methylation. This global demethylation typically results in cells that are unable to survive \cite{shoemaker_mediators_2011}. 

The mechanics of active mechanisms of demethylation is still not well understood. If active demethylation acts to leave a site hemimethylated than we expect to find opposite polarity hemimethylated sites in the locus to be a key signature. Additionally we expect that since active mechanisms don't conserve carbon-carbon bonds, there will not be maintenance of grandmother methylated strands, asymmetric cell division won't be as strong an effect, and the population will be more homogeneous having narrower distributions than those in Fig.~\ref{fig:distributions}(A). 

The model in this paper provides a tool for measuring passive demethylation rates in regulatory scenarios. Additionally, this model may predict if active mechanisms are playing a roll, and in the way that a model of genetic drift provides a necessary baseline for measuring the effects of selection, this model can be used in elucidating mechanisms of active demethylation.

\bibliographystyle{model3-num-names}
\bibliography{MethExhaustion}

\end{document}